\documentclass[3p,times,twocolumn]{elsarticle}

\usepackage{ecrc}

\volume{00}

\firstpage{1}

\journalname{Nuclear Physics B Proceedings Supplement}

\runauth{J.~Gonzalez--Fraile}

\jid{nuphbp}

\jnltitlelogo{Nuclear Physics B Proceedings Supplement}

\usepackage{amssymb}

\biboptions{compress}

\usepackage[figuresright]{rotating}

\newcommand{\ie} {{\it i.e.}}

\begin{document}

\begin{frontmatter}



\dochead{}

\title{Effective Lagrangian approach to the EWSB sector}


\author{J.~Gonzalez--Fraile}
\ead{fraile@thphys.uni-heidelberg.de}

\address{Departament d'Estructura i Constituents de la Mat\`eria and
  ICC-UB, Universitat de Barcelona.
  \\
Institut f\"{u}r Theoretische Physik, Universit\"{a}t Heidelberg.
}

\begin{abstract}
In a model independent framework, the effects of new physics at the electroweak scale can be parametrized in terms of an effective Lagrangian
expansion. Assuming the $SU(2)_L x U(1)_Y$ gauge symmetry is linearly realized, the expansion at the lowest order span dimension--six operators
built from the observed Standard model (SM) particles, in addition to a light scalar doublet. After a proper choice of the operator basis we present a
global fit to all the updated available data related to the electroweak symmetry breaking sector: triple gauge boson vertex (TGV) collider
measurements, electroweak precision tests and Higgs searches. In this framework modifications of the interactions of the Higgs field to the
electroweak gauge bosons are related to anomalous TGV's, and given the current experimental precision, we show that the analysis of the latest
Higgs boson data at the LHC and Tevatron gives rise to strong bounds on TGV's that are complementary to those from direct TGV measurements.
Interestingly, we present how this correlated pattern of deviations from the SM predictions could be different for theories based on
a non--linear realization of the $SU(2)_L x U(1)_Y$ symmetry, characteristic of for instance composite Higgs models. Furthermore, anomalous TGV signals
expected at first order in the non--linear realization may appear only at higher orders of the linear one, and viceversa. Their study could lead to hints
on the nature of the observed boson.
\end{abstract}


\end{frontmatter}


\section{Introduction}
\label{sec:intro}

After the discovery of the Higgs boson~\cite{discovery}, we can finally analyze a
particle that seems directly related to the electroweak symmetry breaking (EWSB) mechanism.
Thus, almost fifty years since the Standard model (SM) Higgs boson was postulated~\cite{higgsteo}, the study
of the discovered particle properties can be used for the first time as a way to
access the mechanism responsible for the EWSB. In particular, in the present note we focus
on the analysis of the Higgs interactions to the rest of SM particles.
A huge variety of data has already been collected, not only from
Higgs searches at both Tevatron and the LHC,
but also from the measurements of triple gauge boson
vertices (TGV)'s at the colliders
and from the low energy electroweak precision measurements. In this context
we present a framework suitable to study the nature of the observed state via the analysis
of its couplings, exploiting all the existing data sets. Pursuing a model independent analysis of the Higgs
interactions, the effective Lagrangian approach~\cite{effective1,effective2,effective3} is arguably
one of the most motivated frameworks from the theoretical point of view, given the lack of observation
of any other new resonance. Hence, the effective Lagrangian is useful as a model independent
way to parametrize the low energy deviations in the Higgs interactions caused by New Physics (NP).
Following this motivation we present on the first part of this note the effective
Lagrangian expansion based on the assumption that the observed state is introduced as a doublet of
$SU(2)_L$, where then the $SU(2)_L \otimes U(1)_Y$ symmetry is linearly realized in the effective
theory which describes the indirect NP effects at LHC
energies~\cite{linearlag}. After presenting the results from a global
analysis to the different EWSB related data sets~\cite{constraining,power}, we focus
on the very interesting complementarity between Higgs analyses and TGV measurements at
colliders~\cite{determining}. On the second part we compare this phenomenology with the one that is
derived from abandoning the assumption of the Higgs as an $SU(2)_L$ doublet. We
study in this case the Higgs interactions using the non--linear or chiral effective Lagrangian including a dynamical Higgs
boson~\cite{Alonso:2012px}. We observe that the interesting Higgs to gauge boson coupling correlation with the TGV
interactions could be lost in this case, besides higher order differences, that both could lead to
phenomenological observable consequences~\cite{Brivio:2013pma} useful to disentangle the nature of
the observed state.

\section{Lagrangian for an elementary Higgs}
\label{sec:linear}

We start assuming that, even if there is NP associated with the EWSB sector, 
the observed particle is an elementary state which belongs to a light electroweak doublet scalar,
and consequently, that the $SU(2)_L \otimes U(1)_Y$ symmetry is linearly realized in the effective
theory. Thus we think of the SM as an effective low energy theory but we still
retain the gauge group, the particle spectrum and the pattern of
spontaneous symmetry breaking as valid ingredients to describe Nature
at energies $E\ll\Lambda$, where $\Lambda$ is the scale associated to NP. This fixes the complete set
of higher dimensional operators that need to be considered at a given order.
Neglecting the effects of the dimension--five total lepton number
violating operator, the lowest order effective operators that can be
built are of dimension--six:
\begin{equation}
\hspace*{1.8cm}
{\cal L}_{\rm eff} =  \sum_n \frac{f_n}{\Lambda^2} {\cal O}_n \;\; ,
\label{l:eff}
\end{equation}
where the operators ${\cal O}_n$ have couplings $f_n$.
If we restrict to $C$ and $P$--even dimension--six operators, 29 of them are relevant to study the Higgs
interactions~\cite{power} barring flavor structure and Hermitian
conjugations. Eight of those modify the Higgs interactions to the
electroweak gauge bosons, one to gluons and one affects only Higgs
self interactions. Additionally there are three Yukawa--like fermionic operators, and
the remaining ones modify both the fermionic couplings to the Higgs boson as well as
to the gauge bosons. Finally TGV interactions of on-shell $W$'s are modified by four of
these operators, as well as, by one operator 
that only involves the electroweak gauge--boson self--couplings, 
${\cal O}_{WWW}$, see~\cite{determining}.

Previous to the analysis of the Higgs data the equations of motion can be used to
eliminate redundant operators from ${\cal L}_{\rm eff}$. In addition,
several operators are already strongly constrained by the use of electroweak precision
data (EWPD), to which they contribute at the tree level. For a detailed discussion on the reduction
of the number of parameters in our
effective Lagrangian see~\cite{power}. After the choice of the basis presented there,
the effective Lagrangian relevant for the analysis of Higgs and TGV data, and 
the subdominant effects to EWPD is reduced to
\begin{eqnarray}
{\cal L}_{\rm eff} \!\!\!
&=& \!\!- \frac{\alpha_s v}{8 \pi} \frac{f_g}{\Lambda^2} 
{\cal O}_{GG}
+ \frac{f_{WW}}{\Lambda^2} {\cal O}_{WW}
+ \frac{f_{BB}}{\Lambda^2} {\cal O}_{BB}\nonumber\\
&+& \frac{f_{\Phi,2}}{\Lambda^2} {\cal O}_{\Phi,2}
+ \frac{f_{\rm bot}}{\Lambda^2}  {\cal O}_{d\Phi,33} 
+ \frac{f_{\tau}}{\Lambda^2} {\cal O}_{e\Phi,33} \label{ourleff}\\
&+& \frac{f_{W}}{\Lambda^2} {\cal O}_{W}
+ \frac{f_{B}}{\Lambda^2} {\cal O}_{B}
\;\; \nonumber
\end{eqnarray}
with 
\begin{eqnarray}
&&\hspace*{-1.0cm}{\cal O}_{GG} = \Phi^\dagger \Phi \; G_{\mu\nu}^a G^{a\mu\nu}  \;,
\;\;
{\cal O}_{WW} = \Phi^{\dagger} \hat{W}_{\mu \nu} 
 \hat{W}^{\mu \nu} \Phi  \; , \nonumber \\
&&\hspace*{-1.0cm} {\cal O}_{BB} = \Phi^{\dagger} \hat{B}_{\mu \nu} 
  \hat{B}^{\mu \nu} \Phi , \;\; 
  {\cal O}_{\Phi,2} = \frac{1}{2} 
\partial^\mu\left ( \Phi^\dagger \Phi \right)
\partial_\mu\left ( \Phi^\dagger \Phi \right) \; , \nonumber \\
&&\hspace*{-1.0cm}{\cal O}_{e\Phi,ij}=(\Phi^\dagger\Phi)(\bar L_{i} \Phi e_{R_j}) 
\; ,  \;\;
{\cal O}_{d\Phi,ij}
=(\Phi^\dagger\Phi)(\bar Q_{i} \Phi d_{Rj})\; , \nonumber  \\
&&\hspace*{-1.0cm}{\cal O}_W  = (D_{\mu} \Phi)^{\dagger}  
 \hat{W}^{\mu \nu}  (D_{\nu} \Phi) \; , \;\;
 {\cal O}_B  =  (D_{\mu} \Phi)^{\dagger} 
  \hat{B}^{\mu \nu}  (D_{\nu} \Phi)  \; ,
\hspace*{0.6cm}
\label{ourope}
\end{eqnarray}
where $\Phi$ is the Higgs doublet, $D_\mu\Phi= \left(\partial_\mu+i \frac{1}{2} g' B_\mu +
i g \frac{\sigma_a}{2} W^a_\mu \right)\Phi$ and $v$ is the vacuum expectation
value. $\hat{B}_{\mu \nu} = i \frac{g'}{2} B_{\mu \nu}$ and $\hat{W}_{\mu\nu} = i \frac{g}{2}
\sigma^a W^a_{\mu\nu}$ with $SU(2)_L$ ($U(1)_Y$) gauge coupling $g$ ($g^\prime$) and Pauli matrices 
$\sigma^a$.

The dimension--six operators in the final basis to be studied, Eq.~(\ref{ourleff}),
contribute to the interactions of the Higgs
boson with the SM particles, in some cases adding new Lorentz structures in the different vertices,
as it was discussed in detail in~\cite{power}. In addition, two of them, ${\cal O}_W$ and ${\cal O}_B$,
contribute to the TGV interactions $\gamma W^+ W^-$ and $Z W^+W^-$, whose contributions
in the commmonly used parametrization of~\cite{Hagiwara:1986vm} are:
%
%
\begin{figure*}[htb!]
  \centering
  \includegraphics[width=0.9\textwidth]{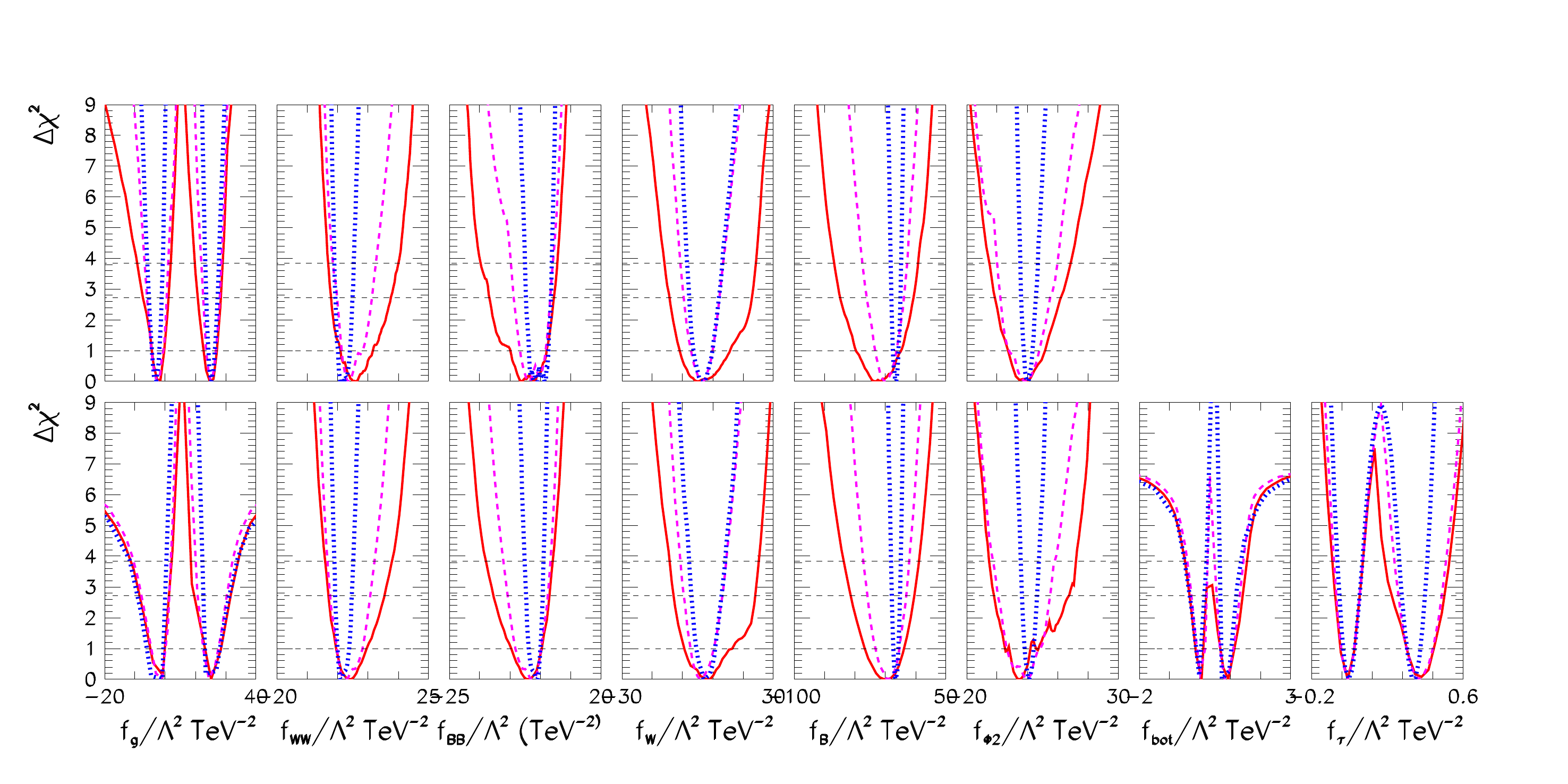}
  \caption{$\Delta\chi^2$ dependence on the fit parameters when we
consider all Higgs collider data (red solid lines), Higgs collider plus TGV data (dashed purple lines), and
Higgs collider plus TGV and EWPD (dotted blue lines). The columns
depict the $\Delta\chi^2$ dependence with respect to the fit
parameter shown in the bottom, with the anomalous couplings $f/\Lambda^2$ given in TeV$^{-2}$, while the rest of undisplayed
parameters are marginalized. In the upper row we use $f_g$, $f_{WW}$, $f_{BB}$, $f_W$, $f_B$, and $f_{\Phi,2}$
as fit parameters with $f_{\rm bot} = f_\tau =0$, while in the bottom row the fitting parameters are $f_g$, $f_{WW}=-f_{BB}$, $f_W$,
$f_B$, $f_{\Phi,2}$, $f_{\rm bot}$ and $f_\tau$.}
\label{fig:1dima}
\end{figure*}
%
%
\begin{eqnarray}
&&\hspace{0.3cm}\Delta \kappa_\gamma = 
 \frac{g^2 v^2}{8\Lambda^2}
\Big(f_W + f_B\Big)\,,
\nonumber \\  
&&\hspace{-1.0cm}\Delta g_1^Z= \frac{g^2 v^2}{8 c^2\Lambda^2}f_W \, ,\; 
\Delta \kappa_Z =   \frac{g^2 v^2}{8 c^2\Lambda^2}
  \Big(c^2 f_W - s^2 f_B\Big)\, ,\hspace*{0.5cm}
\label{eq:wwv}
\end{eqnarray}
with $\kappa_V=1+\Delta \kappa_V$ and $g^Z_1=1+\Delta g^Z_1$, and where
$s$($c$) is the sine (cosine) of the weak mixing angle. For completeness
we note that ${\cal O}_{WWW}$ contributes to the TGV
parametrization in~\cite{Hagiwara:1986vm} as
$\lambda_\gamma = \lambda_Z = \frac{3 g^2 M_W^2}{2 \Lambda^2} f_{WWW}$,
although it has no relevance for the present note. Finally, ${\cal O}_{W}$,
${\cal O}_{B}$, ${\cal O}_{WW}$, ${\cal O}_{BB}$ and ${\cal O}_{\phi,2}$
also give subdominant (loop) contributions to EWPD.

In order to constrain these higher dimensional deviations
with respect to the Standard Model expected behaviour we build a $\chi^2$ function based on the signal
strength measurements of the Higgs production and decay at both the Tevatron and LHC, together
with the addition of the most precise determination of TGV's at the colliders in the present framework,
as well as the inclusion of EWPD in a simplified way in terms of the oblique parameters
$S,\ T$ and $U$. The details on the data included on this global analysis, as well as the statistical
details of the fit are described thoroughly in~\cite{constraining,power}.

\subsection{Results}

In order to study the present bounds on the Higgs interactions from Higgs, TGV and EWPD sets we perform
several global analyses including different data and dimension--six operator sets. The results are shown
in Fig.~\ref{fig:1dima}.

In the Figure one can see the $\Delta\chi^2$ dependence on the fit parameters when we
consider all Higgs collider data (red solid lines), Higgs collider plus TGV data (dashed purple lines), and
Higgs collider plus TGV and EWPD (dotted blue lines). In the columns we show the $\Delta\chi^2$ dependence with respect to the fit
parameter written in the bottom of the column, with $f/\Lambda^2$ given in TeV$^{-2}$, and while the rest of undisplayed
parameters have been marginalized. In the upper row we use $f_g$, $f_{WW}$, $f_{BB}$, $f_W$, $f_B$, and $f_{\Phi,2}$
as fit parameters with $f_{\rm bot} = f_\tau =0$, while in the bottom row the fitting parameters are $f_g$, $f_{WW}=-f_{BB}$, $f_W$,
$f_B$, $f_{\Phi,2}$, $f_{\rm bot}$ and $f_\tau$. With the precision reachable from the considered data sets the main conclusion that
we can extract is that so far all the considered Higgs and gauge interactions are completely compatible with the SM Higgs hypothesis.
This causes all the $\Delta\chi^2$ panels to have the minima close to the 0 point (SM point) without any statistically significant deviation
with respect to the SM pattern of interactions. The degeneracies in $f_g$, $f_{\rm bot}$ and $f_\tau$ are due to the interference with the
SM corresponding vertices. In addition, by comparing the results for the upper and lower operator sets considered, we can observe that the
introduction of the fermionic operators to the analysis has a negligible effect on ${\cal O}_{W}$,
${\cal O}_{B}$, ${\cal O}_{WW}$, ${\cal O}_{BB}$ and ${\cal O}_{\phi,2}$. An exception is the observed weakening of the exclusion ranges
on $f_g$, which is
strongly related to the inclusion of $f_{\rm bot}$, as discussed in~\cite{power}. Fig.~\ref{fig:1dima} is useful to illustrate some of the interesting
features and the potential of the data--driven approach, as it allows us to identify which of the data sets impose the strongest constraints on
a given operator. For instance, by looking at either the $f_W$ or $f_B$ panels, either the upper or the lower row, we could
already foresee comparing the results from the analysis of the Higgs data to the results from the combined analysis of Higgs and TGV data
that the precision reachable from both data sets on $f_W$ or $f_B$ is at a similar level. A conclusion that drives us directly to the following
subsection.

\subsection{Determining TGV from Higgs data}

\begin{figure}
  \centering
  \includegraphics[width=0.45\textwidth]{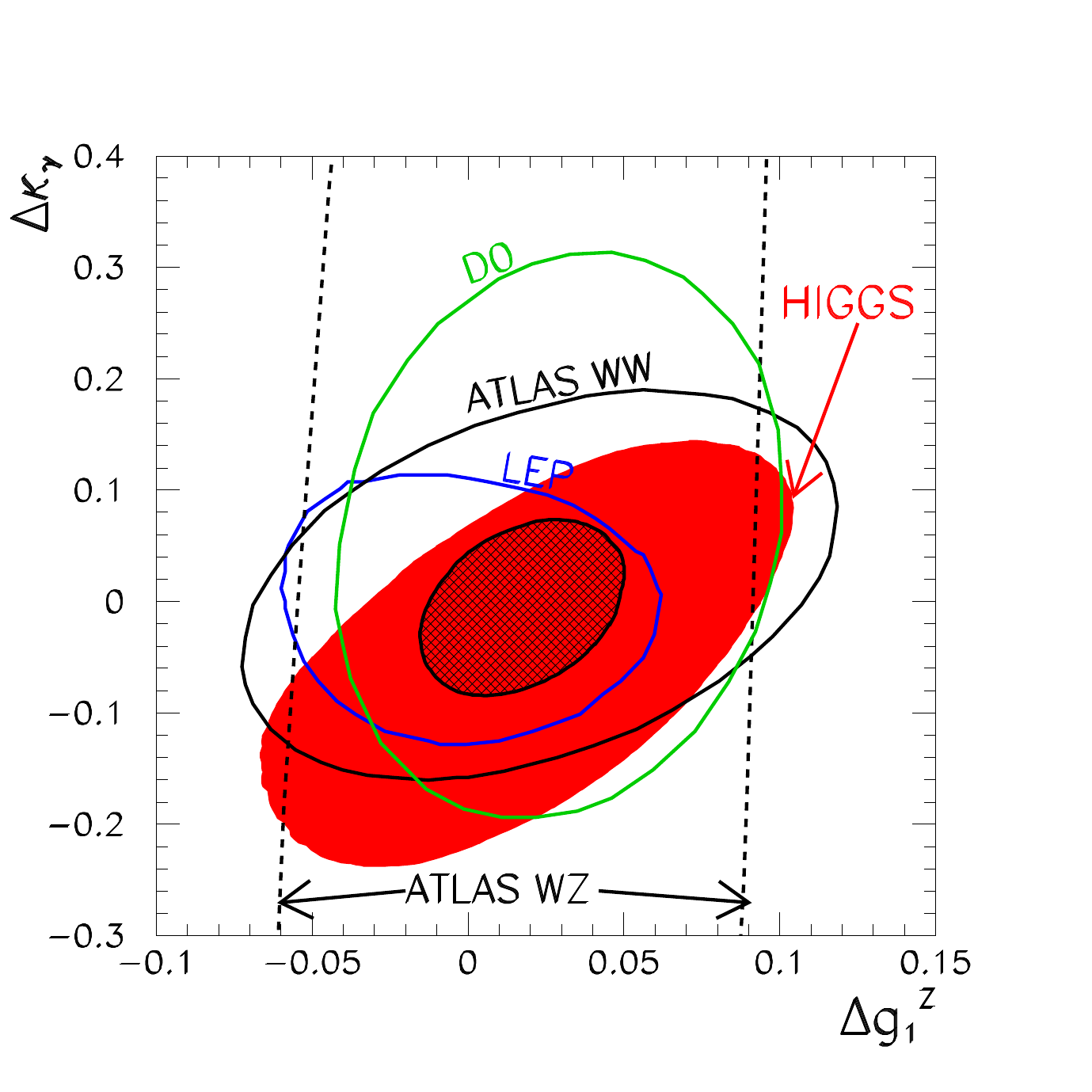}
 \caption{95\% C.L. allowed regions (2 d.o.f.) on the plane
   $\Delta\kappa_\gamma \otimes \Delta g^Z_1$ from the global analysis of only
   Higgs data from the LHC and Tevatron (red filled region), and relevant bounds from
   TGV measurements in different collider experiments as labeled in the panel. We also
   show the estimated constraints obtainable by combining these bounds (hatched region).}
\label{fig:tgv}
\end{figure}

As we have commented, two of the operators, ${\cal O}_W$ and ${\cal O}_B$, contribute both to the Higgs interactions with the gauge
bosons, and to TGV interactions. In the results presented in Fig.~\ref{fig:1dima} we have used this double contribution on the
direction of adding TGV data to the global analysis of the dimension--six operators in order to further constrain the coefficients
of the operators
considered in Eq.~(\ref{ourleff}). Conversely, we can exploit this double contribution on qualitatively the opposite direction. We present
now the global analysis of only Higgs data, but this time we rotate the exclusion bounds derived from the results to bounds on the
TGV parameters by using
Eq.~(\ref{eq:wwv}). As we have observed in the previous subsection that the fermionic operators do not affect the electroweak ones, for
simplicity we show in this subsection the
results of the space spanned by $f_g, f_{WW}, f_{BB}, f_{\Phi,2}, f_W, \mbox{and } f_B$. We present the results of the analysis to the Higgs data
only in Fig.~\ref{fig:tgv}. There we plot in the red filled region the 95\% C.L. allowed values in the plane
$\Delta\kappa_\gamma \otimes \Delta g_1^Z$ after marginalizing over the other four parameters on the Higgs analysis, $f_g,f_{WW},f_{BB}$ and
$f_{\Phi,2}$, \ie\ we define
\begin{eqnarray}
&&\hspace*{-1.5cm}\Delta\chi^2_{H}(\Delta\kappa_\gamma,\Delta g^Z_1) 
=\\
&&\hspace*{-1.0cm}\ {\rm min}_{f_g,f_{WW},f_{BB},f_{\Phi,2}}
\Delta\chi^2_H(f_g,f_{WW},f_{BB},f_{\Phi,2},f_B,f_W) \;\; ,
\nonumber
\end{eqnarray}
and we impose the two-dimensional 95\% C.L. allowed region from the condition
$\Delta\chi^2_{H}(\Delta\kappa_\gamma,\Delta g^Z_1)\leq 5.99$.
In the Figure we also show the relevant 95\% C.L. bounds from TGV measurements in different collider experiments
as properly labeled. These experimental measurements were all performed in the framework given by Eq.~(\ref{eq:wwv}),
for further details see~\cite{determining}. As we can observe in the Figure, the direct measurements
of TGV's and the results derived from the global analysis of Higgs data translated to the TGV space, are not only
complementary because of the different correlation that they present, but actually the precision reachable from
both types of determinations are at a comparable level. As a consequence, this interesting complementarity between both
analysis can be used as a further test of the SM and of the linearly realized EWSB. In the future,
when the data sets are extended, the combination of both types of analysis has the potential to furnish the strongest possible
bounds on this anomalous TGV space. Indeed, after performing the current combination of results, as described
in~\cite{determining}, the hatched region in Fig.~\ref{fig:tgv} sets the strongest 95\% C.L. exclusion region of the ones
currently available.

\section{Disentangling a dynamical Higgs}

\begin{figure}[ht!]
\centering
\includegraphics[width=0.41\textwidth]{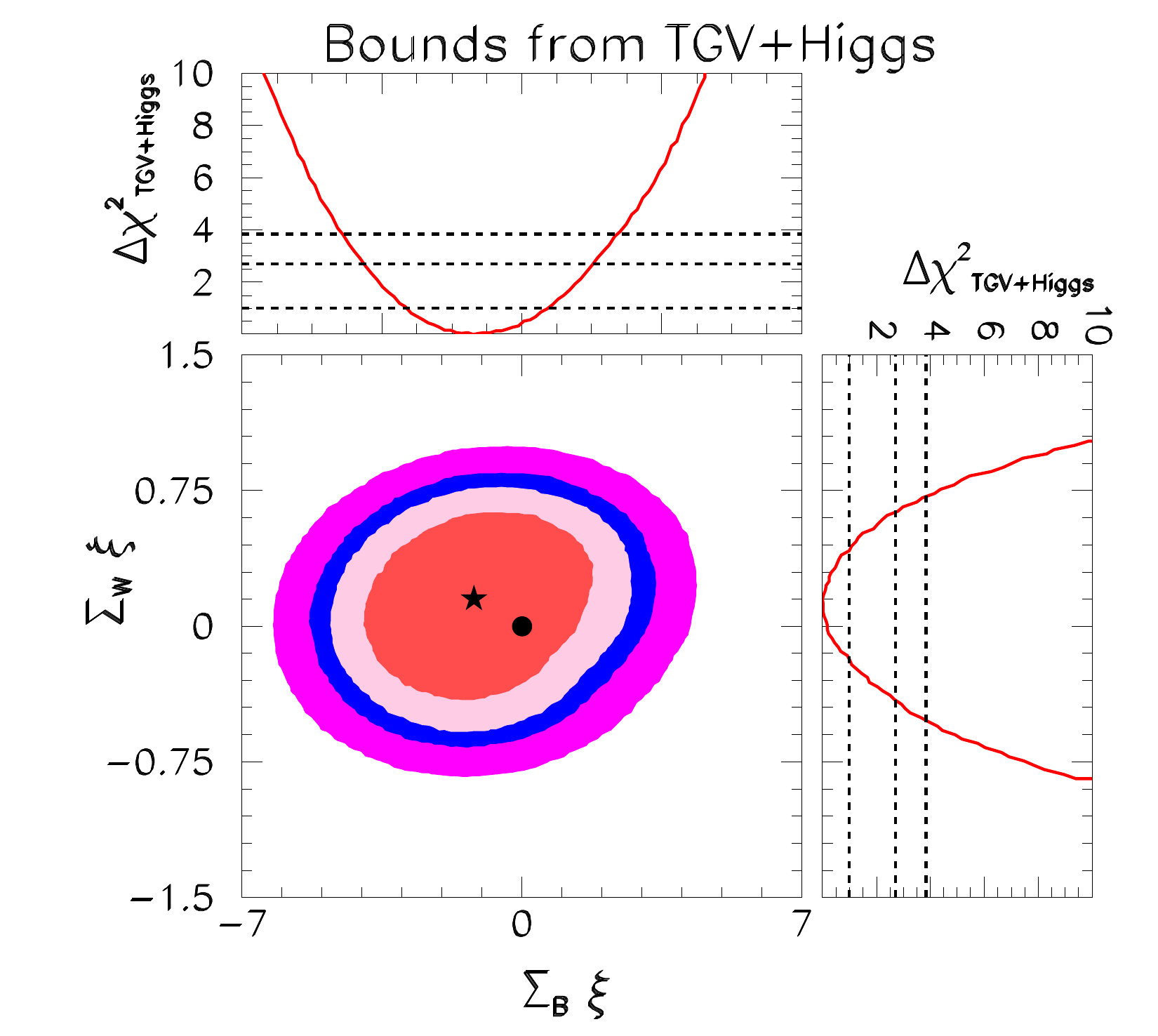}
\includegraphics[width=0.41\textwidth]{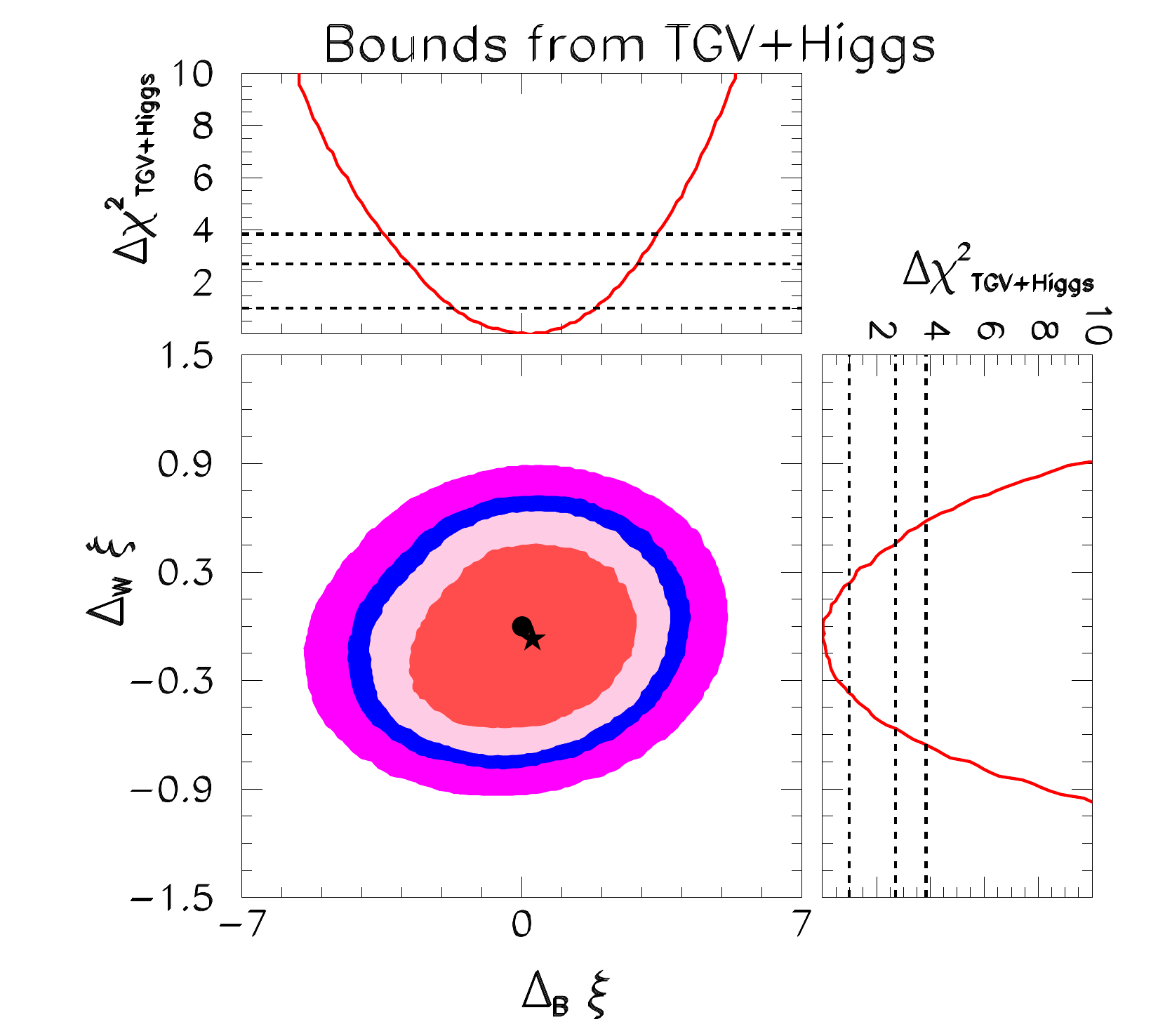}
\caption{\em
{\bf Upper}: A BSM sensor irrespective of the type of expansion:
  constraints from TGV and Higgs data on the combinations
  $\Sigma_B=4(2c_2+a_4)$ and $\Sigma_W=2(2c_3-a_5)$, which converge to
  $f_B$ and $f_W$ in the linear $d=6$ limit. 
The dot  at $(0,0)$ signals the SM expectation.
{\bf Lower}: A non-linear versus linear discriminator: constraints on the
combinations $\Delta_B=4(2c_2-a_4)$ and $\Delta_W=2(2c_3+a_5)$, which
would take zero values in the linear (order $d=6$) limit (as well
as in the SM), indicated by the dot at $(0,0)$.  
For both figures the lower left panels
  shows the 2-dimensional allowed regions at 68\%, 90\%, 95\%, and
  99\% CL  after marginalization with respect to the rest of parameters
  spanned in the analysis, see~\cite{Brivio:2013pma}.
The star corresponds to the best fit point of the analysis. 
The upper left and lower right panels in each figure give the
corresponding 1-dimensional projections over each of the two
combinations considered.
}
\label{fig:nl1}
\end{figure}

We proceed now to compare the phenomenology of the effective Lagrangian expasion assuming a linear realization
of the EWSB with the phenomenology of the non--linear or chiral Lagrangian expansion, including a light Higgs, but in this case
generically as a singlet of $SU(2)_L$.
While the first expansion is suitable for models with elementary Higgs particles, the second one may be more adequated
for ``dynamical" -composite- ones. For the non--linear case we consider the construction of the effective Lagrangian
presented in~\cite{Alonso:2012px}, up to four
derivatives in the expansion and including the bosonic and Yukawa--like $C$ and $P$ even
operators\footnote{With an increased precision
some of these assumptions may be relaxed, allowing for
instance for $C$ and $P$ violating operators to be included and
studied~\cite{Gavela:2014vra}.}.
For the comparison of the phenomenology, we can start restricting to the subset of chiral operators that include the same
gauge interactions that are already introduced by the dimension--six operators in the linear expansion, \ie\ we consider
the chiral structures weighted by the parameter $\xi$ in~\cite{Alonso:2012px,Brivio:2013pma}.
The lesser symmetry constraints in the non--linear effective Lagrangian means more possible invariant
operators at a given order, which is translated into phenomenological decorrelations with respect to the linear case, differences
that we present here. For instance, in the unitary gauge, two of the non--linear operators, ${\cal P}_2(h)$ and ${\cal P}_4(h)$
in the notation of~\cite{Brivio:2013pma}, read:
\begin{small}
\begin{eqnarray}
&&\hspace*{-1.2cm}{\cal P}_2(h) =2 ieg^2A_{\mu\nu}W^{-\mu} W^{+\nu}{\cal F}_2(h) - 2 \frac{ie^2g
}{\cos\theta_W} Z_{\mu\nu}W^{-\mu} W^{+\nu}{\cal F}_2(h)\,,
\label{P2unitary}\hspace*{0.4cm}\\
&&\hspace*{-1.2cm}{\cal P}_4(h)= - \frac{eg}{\cos\theta_W} A_{\mu\nu}Z^\mu\partial^\nu
{\cal F}_4(h) + \frac{e^2}{ \cos^2\theta_W}
Z_{\mu\nu}Z^\mu\partial^\nu{\cal F}_4(h)\,,\hspace*{0.4cm}
\label{P4unitary}
\end{eqnarray}
\end{small}
with their coefficients $c_2$ and $c_4$ taking arbitrary
(model dependent) values, and where the model dependent functions ${\cal F}(h)$ encode the
chiral interactions of the light $h$. In contrast, the $d=6$ operator containing the same gauge
interactions, ${\cal O}_B$, results in the combination:
\begin{small}
\begin{eqnarray}
\hspace*{-0.8cm}{\cal O}_B =&\hspace*{-0.3cm} \frac{ieg^2}{8} A_{\mu\nu}W^{-\mu} W^{+\nu}(v+h)^2
-\frac{ie^2g}{8\cos\theta_W} Z_{\mu\nu}W^{-\mu} W^{+\nu}(v+h)^2\hspace*{0.2cm} \\
&\hspace*{-0.3cm} -\frac{eg}{4\cos\theta_W} A_{\mu\nu}Z^\mu\partial^\nu
h(v+h)+\frac{e^2}{4\cos^2\theta_W} Z_{\mu\nu}Z^\mu\partial^\nu h(v+h)\,.\hspace*{0.2cm}
\label{OBunitary}
\end{eqnarray}
\end{small}
In consequence, in a general non--linear analysis the $\gamma WW$ interactions get decorrelated from
the $\gamma Zh$ and $ZZh$ couplings, all encoded in ${\cal O}_B$, and for the precise Lorentz structures shown above.
These interactions construct in the linear case the correlation that leads to the interesting complementarity between
Higgs data analysis and the direct measurement of TGV's that we have highlighted in the previous section when exploiting the
effective Lagrangian analysis in the linear realization of the EWSB. Thus
in order to study whether the EWSB mechanism presents a correlated or a decorrelated behaviour, we can perform
the global analysis of Higgs, TGV and EWPD in the context of the non--linear effective Lagrangian, what is the first
8--dimensional analysis in this basis~\cite{Brivio:2013pma}. While the complete details of such study can be
found in~\cite{Brivio:2013pma}, here we only show a subset of the two--dimensional results. But instead of showing the
ranges of results as a function of the coefficients of the non--linear operators included in the global analysis,
we rotate part of the coefficients to a new set of variables, see~\cite{Brivio:2013pma}. This is illustrated in
Fig.~\ref{fig:nl1}, where the results of the combined analysis of the data sets considered have been projected into
combinations of the non--linear operators ${\cal P}_2(h)$, ${\cal P}_3(h)$, ${\cal P}_4(h)$ and ${\cal P}_5(h)$: 
\begin{eqnarray}
&&\hspace*{-1.2cm} \Sigma_B\equiv 4(2c_2+a_4)\,, \qquad\qquad \Sigma_W\equiv 2(2c_3-a_5)\,,\\
&&\hspace*{-1.2cm} \Delta_B\equiv 4(2c_2-a_4)\,, \qquad\qquad \Delta_W\equiv 2(2c_3+a_5)\,,
\end{eqnarray}
defined such that at order $d=6$ of the linear expansion $\Sigma_B=f_B$, $\Sigma_W=f_W$, while $\Delta_B=\Delta_W=0$.
In the expressions $c_i$ and $a_i$ stand for the relevant coefficients of the non--linear operators considered for the
comparison~\cite{Brivio:2013pma}, where in the non--linear case we have included the light Higgs in the generic functions
in Eqs.~(\ref{P2unitary}) and~(\ref{P4unitary}), assuming
${\cal F}_i(h)\equiv1+2\tilde a_i\frac{h}{v}+\ldots$,
where the dots stand for higher powers of $h/v$ that are irrelevant for the present note, and where
to further simplify the notation $a_i\equiv c_i\tilde a_i $, with $c_i$ being the global chiral
operator coefficients.
With these variables, the $(0,0)$ coordinate corresponds to the SM in the upper figure of
Fig.~\ref{fig:nl1}, while in the lower figure it corresponds to the linear regime (at order $d=6$). Would future
data point to a departure from $(0,0)$ in the variables of the first
figure it would indicate BSM physics irrespective of the linear or
non-linear character of the underlying dynamics; while such a
departure in the bottom one would be consistent with a non-linear
realization of EWSB.

Additional phenomenological differences between both effective Lagrangian expansions
were further studied in~\cite{Brivio:2013pma}. They involve interactions that may be generated
by leading operators in one of the expansions, while they can only be generated by subleading--
and then suppressed-- operators in the alternative approach. Several examples of this type
of higher order differences, as well as a realistic analysis of the LHC capability to improve the precision
on some of the interactions involved can be found in~\cite{Brivio:2013pma}. They require a more
detailed analysis of anomalous TGV's.

\section{Conclusions}

In this note we have presented some of the interesting features of the model independent analysis of the
Higgs interactions by means of an effective Lagrangian expansion. In the first part we have presented
a global analysis of Higgs, TGV and EWPD assuming that the Higgs is part of an $SU(2)_L$, and thus that
the EWSB is linearly realized. From this analysis in terms of dimension--six operators we have concluded
that currently the considered Higgs couplings to the SM particles as well as the studied gauge self--interactions
are completely compatible
with the SM Higgs hypothesis within the reachable precision. With the data--driven approach that we have followed we have been able to
indentify a very interesting complementarity between the direct measurement at the colliders of the TGV
interactions and the results obtained from a global analysis of the Higgs data. The combination of both
types of analysis has the potential to lead to the strongest sensitivity on deviations from the SM pattern
of interactions in the TGV's, being useful as a consequence to further test the SM and the linearly realized EWSB.
Interestengly we have illustrated that this correlated pattern of interactions may disappear when we consider
an alternative expansion, the non--linear effective Lagrangian. In this case the lesser symmetry constraints
could be translated in general scenarios in the decorrelation between Higgs to gauge bosons and TGV interactions.
We have presented a possible set of variables to study how the collected data allow us to disentangle
the two possible expansions, and as a consequence the nature of the observed state and EWSB mechanism.\\

The contents of this note are based on the published works~\cite{constraining,power,determining,Brivio:2013pma},
the author thanks his collaborators in these studies. J.G-F acknowledges support from
European Union network FP7 ITN INVISIBLES (Marie Curie Actions, PITN-GA-2011-289442),
MICINN FPA2010-20807, consolider-ingenio 2010 program CSD-2008-0037 and ME FPU grant AP2009-2546.




\nocite{*}
\bibliographystyle{elsarticle-num}



\end{document}